# Characteristics of neutrons and proton beams arising from two different Beam Nozzles

**Yeon-Gyeong Choi and Yu-Seok Kim[*]**

*Department of Energy & Environment Engineering, Dongguk University, Gyeongju 780-714*

Tandem or Van de Graaff accelerator with an energy of 3-MeV is typically used for PIXE analysis. In this study, the beam line design used in PIXE analysis was used to increase the production of isotopes instead of the typical low-energy accelerator from a 13-MeV cyclotron. For PIXE analysis, the proton beam should be focused at the target through the use of a nozzle after degrading the proton beam energy from 13-MeV to 3-MeV using an energy degrader. Previous studies have been conducted to determine the most appropriate material and thickness of the energy degrader. Based on the energy distribution of the degraded proton beam and the neutron occurrence rate at the degrader an aluminum nozzle of X thickness was determined to be the most appropriate nozzle construction. Neutrons are created by the collision of 3-MeV protons into the nozzle after passage through energy degrader. In addition, a sufficient intensity of proton beam is required for non-destructive analysis of PIXE. Therefore, in order to optimize nozzle design, it is necessary to consider the number of neutrons which arise from the collision of protons inside the nozzle, as well as the track direction of the generated secondary neutrons, with the primary aim of ensuring a sufficient number of protons passing through the nozzle as a direct beam. A number of laboratories are currently conducting research related to the design of nozzles used in accelerator fields, mostly relating to medical fields. In this paper, a comparative analysis was carried out for two typical nozzle shapes in order to minimize losses of protons and generation of secondary neutrons. The neutron occurrence rate and the number of protons after passing through the nozzle were analyzed using a Particle and Heavy Ion Transport code System (PHITS) program, in order to identify the nozzle which generated the strongest proton beam.

Email: unison@dongguk.ac.kr[*]

Fax: +82-54-770-2857[*]

## I. INTRODUCTION

PIXE non-destructive analysis uses a characteristic x-ray, which is generated by irradiating the proton of a specific energy range to matter. For this purpose, proton beams of 3MeV are optimal [1] [2]. The specific design of the PIXE beam line used in this study is based on a 13MeV cyclotron, and has been previously published [3]. In our previous study, the beam line design was composed of a circular carbon plate, which is used as an energy degrader to attenuate the 13MeV band proton beam energy into the 3MeV band, and a nozzle, which uniformly irradiates the beam after passage through the energy degrader. The beam nozzle was designed using aluminum due to both the ease of fabrication as well as the low incidence of neutron generation. The primary design criterion of the nozzle was a nozzle angle of 20 degrees or less, in order to minimize the collision of the beam with the interior of the nozzle.

Two general nozzle shapes are currently used in PIXE. The first gradually reduces the size of the beam [4], having a nozzle angle of 20 degrees or less, and the second reduces the size of the beam initially in the nature of a collimator, and then gradually reduces the size of the beam [5].

Many studies of beam line design for PIXE analysis using an accelerator relate to the energy degrader, collimator, and the specific design of the nozzle. However, there have been no comparative analysis studies of nozzle shapes, and hence, this paper presents a comparative analysis of two different nozzle design types, considering the number of protons passing through the nozzle and the number of neutrons generated by the collision with the nozzle. To achieve this, the Particle and Heavy Ion Transport code System (PHITS) program [5] was used. The PHITS program is a Monte Carlo particle transport code written in Fotran, and the resulting value can be obtained by using the tally, such as those of heat, yield, and product, by entering input values such as the source and geometry. Using the PHITS program, the number of neutrons generated in the nozzle and the number of protons passing through the nozzle were simulated.

## II. SIMULATION

Typically there are two different shapes of particle beam nozzles used for PIXE analysis and medical purposes, as shown in Figure 1.

The exit diameters of each of the two beam nozzles is 1mm. Nozzle no.1 gradually and continuously reduces the size of proton beam to a diameter of 1mm, whereas nozzle no.2 reduces the size of proton beam in two stages. In both nozzles, the 13-MeV protons which are produced by the cyclotron are degraded to 3-MeV upon collision with the energy degrader, and the degraded protons are then applied to irradiate a target after passing through the nozzle.

Many neutrons are generated upon collision of the proton beam with the interior of the beam nozzle. Secondary neutrons can cause radiation exposure to surrounding materials, and shielding against them is difficult. Therefore, the neutron generation rate should be minimized. Most of the secondary neutrons are generated with a trajectory which is the same as that of the proton beam, while many of the secondary neutrons are generated in at different angles, some of which may be in a direction which is directly opposed to that of the beam trajectory. Secondary neutrons, generated in the opposite direction should be minimized. Furthermore, the number of protons exiting the nozzle with a uniform trajectory should be maximized so as to ensure sufficient beam current for PIXE analysis. In order to analyze these properties, the incidence of secondary neutrons and distribution of secondary neutrons according to the angle of their trajectory were simulated for the two nozzles using the PHITS program.

1. Comparative analysis of the incidence of neutrons.

Source input parameters of the PHITS are indicated in Table. 1.

Beam radiuses were equally designed for each design type, and beam dispersions were selected from 1-degrees to 5-degrees in the source input parameter. In order to compare the incidence of neutrons, a sphere with same size surrounding the nozzle was established for each design type, and the numbers of neutrons passing through these spheres were compared. Figure 2. shows the tracks directions of the neutrons generated in the two different nozzles. If [T-Track] tally of PHITS is used, the flux at a specific region can be obtained. In addition, the number of neutrons generated by a single proton are indicated in Table 2.

In Table 2, the incidence of neutrons arising from the two nozzles are shown to be nearly identical. In addition, neutrons generated are indicated in Figure 2. In nozzle no. 2, many neutrons were generated in the opposite direction to that of the proton beam. In this case, it is difficult to shield neutrons, and losses of protons can be caused. Thus, the angle of the generated neutrons must be determined for each nozzle.

Table 3. shows distribution of the generated neutrons according to the angle of their dispersion. The angle referred to throughout this text is based on the direction of the proton beam, with 180-degree representing a direction opposite to that of the beam. It can be seen that the numbers of the neutrons generated at a low angle (In the direction of the beam nozzle) are almost identical for the two nozzle shapes (Table 3). However, a considerably greater number of neutrons were generated in the opposite direction of the beam when nozzle no.2 was applied, as compared to those generated in nozzle no.1.

Angle distribution of generated neutron, generated by colliding with protons and nozzle, was calculated according to equation (2.1) [6].

$$\frac{dN}{d\omega} = \varphi n \left(\frac{d\sigma}{d\omega}\right) dx \qquad (2.1)$$

where ω is the angle between the generated particle and the progress direction of the projectile, φ is initial flux, n is atom density of the target, dσ is the cross section of a single atom, and x is the thickness of the target. The total cross section σ was calculated according to equation (2.2).

$$\sigma = \int_0^{2\pi} \int_0^{\pi} \frac{d\sigma}{d\omega}(\theta)\sin\theta d\theta d\varphi \qquad (2.2)$$

2. Comparative analysis of the number of protons passing through the nozzle.

Upon passage through the energy degrader, a proton beam of 13MeV is attenuated to an energy of 3MeV. The beam is then directed through the nozzle at the target, at which point a portion of the proton beam collides with the nozzle, resulting in a secondary loss of beam strength. A low current beam is required for this analysis, although the current must not be sufficient to generate suitable and practicable figures. The number of protons lost during passage through the nozzle should be reduced as much as possible. The numbers of protons passing exit holes of the two nozzles are simulated using PHITS. Table 3 and Figure 3 show numbers and directions of the protons passing the nozzles, respectively.

In comparison with the number of protons passing through the nozzle, the number of protons passing through nozzle no.1 was much more than that passing through nozzle no.2. Track distribution of the proton beams at the nozzle was indicated in Figure 3.

Figure 3 indicates the track distribution and flux distribution for protons in the xy axes. In Figure 3, the center is shown in red color, and it can be seen that there is a greater degree of proton dispersion by the proton beam passing through nozzle no.1, as compared to that passing through nozzle no.2. The numbers of neurons generated from the collision of a single proton for each nozzle are 5.6397E-06 and 5.6228E-06 for nozzles 1 and 2, respectively. Thus, no significant difference in the amount of neutron generation by these two types of nozzle was identified. In addition, according to the track direction of the generated neutrons, the application of nozzle no.2 type results in the generation of a greater number of back-scattered neutrons. The number of neutrons passing through nozzles no. 1 and no. 2 were determined as 2.6919E-03 and 1.6293E-03, respectively, and thus, there was a greater loss of neutrons from the application of nozzle no. 1 as compared to that of nozzle no. 2. Where a cyclotron of 13MeV is being utilized, the energy should be attenuated to 3MeV so as to reduce the number of protons to a level which is comparable to that arising from a low energy accelerator. In this study, it was determined that the use of nozzle no. 1 is the most appropriate for PIXE analysis utilizing a high energy accelerator, as a lesser amount of neutron occurs, and this is directly related to a lesser amount of proton loss.

## III. Conclusion

In this study, a comparative analysis of the characteristics of two different designs of proton beam nozzle, produced with the same aluminum material, was carried out using the PHITS program. The parameters which were analyzed included the rate of secondary neutron generation, the track distribution of secondary neutrons, and the number of protons which passed through the nozzle. According to PHITS analysis, nozzle no. 1 was found to enable the more efficient control on the neutron block since most secondary neutrons are generated at the front part unlike in other nozzles in different shapes nozzle no. 1 also satisfies the required current condition for nondestructive analyses, such as PIXE/PIGE, due to a sufficient number of protons passing through the nozzle. Therefore, it was concluded that nozzle no. 1 has the most appropriate shape for PIXE analysis. Further studies of neutron shielding are planned for the future, so as to further refine the application of non-destructive analyses.

## ACKNOWLEDGEMENT

This research was supported by the National Nuclear R&D Program through the National the Research Foundation of Korea (NRF) funded by the Ministry of Education, Science and Technology (2010-0018573)

Table. 1. Source input parameter

| Beam radius (mm) | Dispersion (degree) |
|---|---|
| 5 | 1 |
| 5 | 2 |
| 5 | 3 |
| 5 | 4 |
| 5 | 5 |

Table 2. The number of neutrons passing through the sphere. [#of the generated neutrons per proton]

| Initial beam spread angle (degree) | nozzle no. 1 (All units are E-06) | nozzle no. 2 (All units are E-06) |
|---|---|---|
| 1 | 5.7673 | 5.8335 |
| 2 | 5.7265 | 5.6985 |
| 3 | 5.6872 | 5.6325 |
| 4 | 5.6131 | 5.5628 |
| 5 | 5.4044 | 5.3865 |
| average | 5.6397 | 5.6228 |

Table 3. Comparison of the number of neutrons based on the angle of their dispersion.

| Degree | nozzle no. 1 | nozzle no. 2 |
|---|---|---|
| 30° | 1.1029E-06 | 1.0026E-06 |
| 60° | 5.0792E-07 | 1.0026E-06 |
| 90° | 1.4402E-06 | 1.0734E-06 |
| 120° | 1.1436E-06 | 1.8566E-06 |
| 150° | 1.3556E-06 | 1.5912E-06 |
| 180° | 2.0268E-07 | 7.4682E-07 |

Table 4. The number of protons passing through nozzle

| Initial beam spread angle (degree) | nozzle no. 1 | nozzle no. 2 |
|---|---|---|
| 1 | 5.6772E-03 | 3.4876E-03 |
| 2 | 3.7042E-03 | 2.2286E-03 |
| 3 | 2.0857E-03 | 1.2426E-03 |
| 4 | 1.2155E-03 | 7.2239E-04 |
| 5 | 7.7671E-04 | 4.6510E-04 |
| average | 2.6919E-03 | 1.6293E-03 |

Figure Captions.

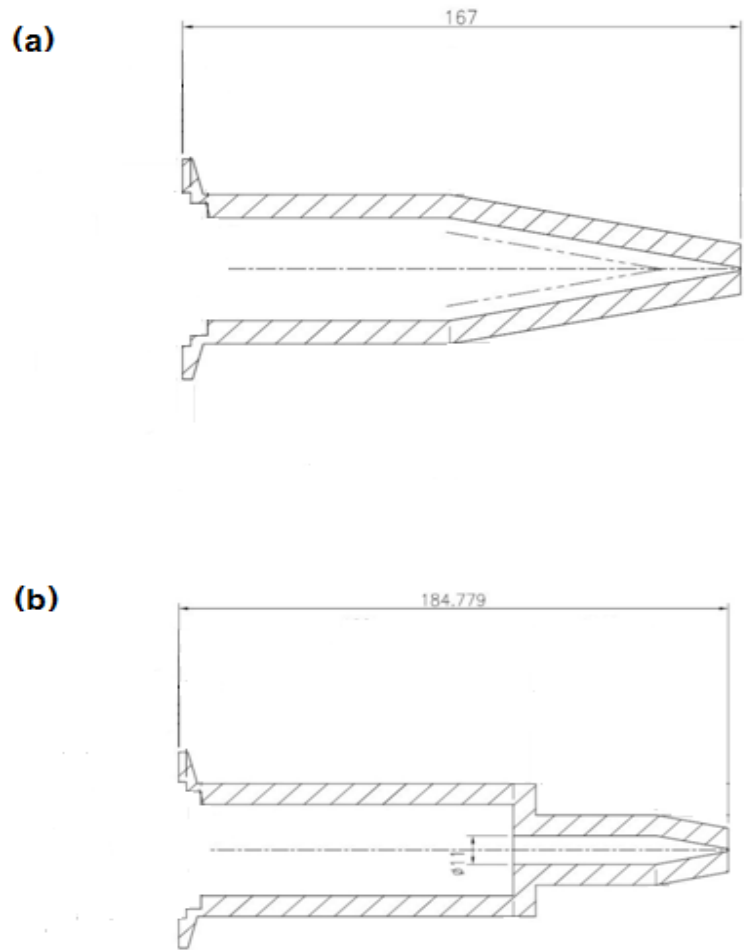

Figure 1. Nozzle shapes used for proton beams (a) nozzle no.1, (b) nozzle no.2

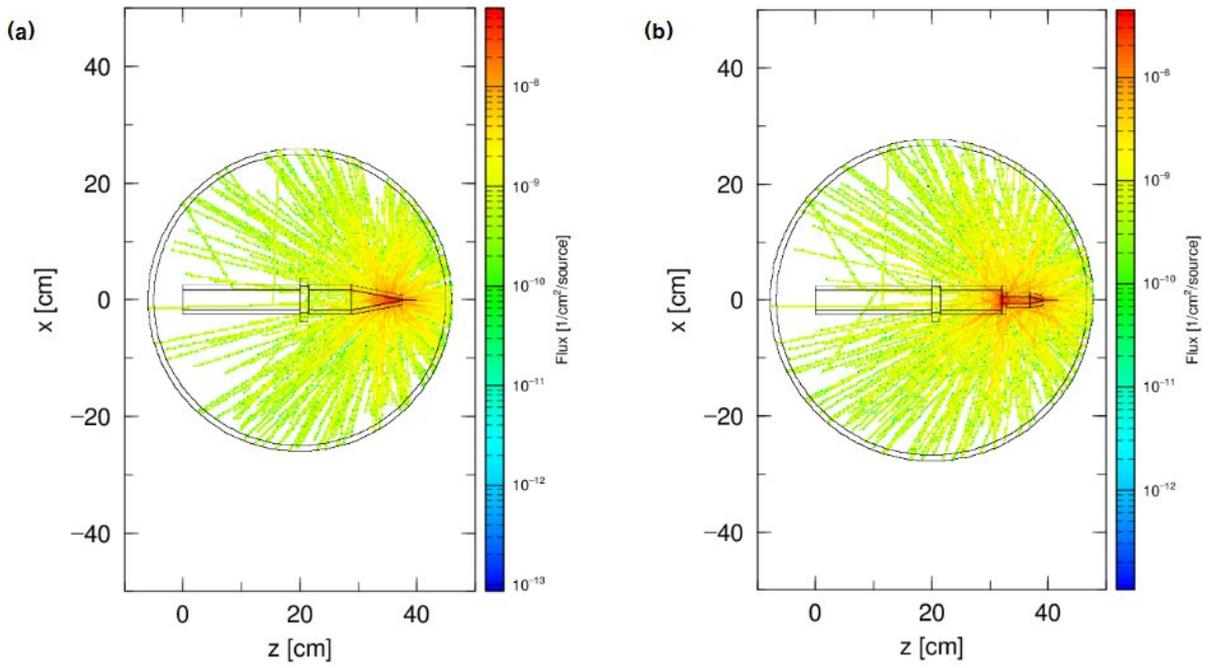

Figure 2. Track distribution of neutrons generated at the nozzle : (a) nozzle no.1, (b) nozzle no.2

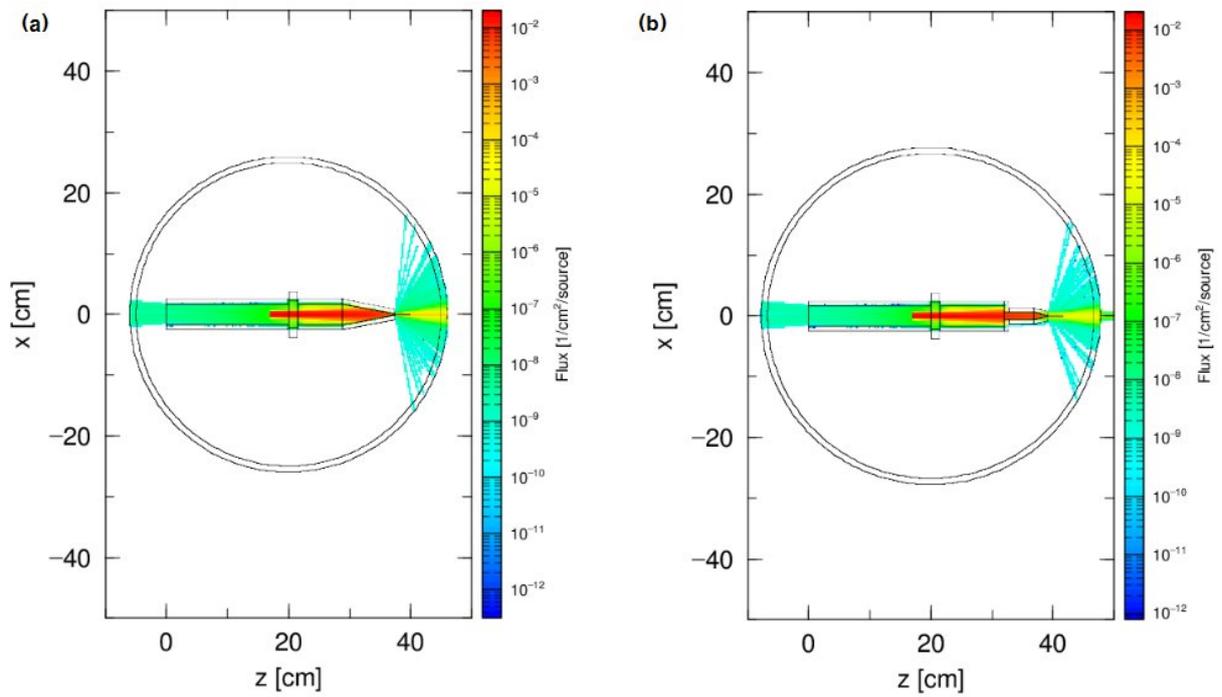

Figure 3. Track distribution of protons at the nozzle : (a) : nozzle no.1, (b) : nozzle no.2